\documentclass{article}

\usepackage{spconf,amsmath,graphicx}
\usepackage[utf8]{inputenc}
\usepackage{amsmath}
\usepackage{algorithm}
\usepackage{algorithmic}
\usepackage{commath}
\usepackage{float}
\usepackage[british]{babel}

\usepackage{cleveref}

\title{\bf Cortical surface parcellation based on intra-subject white matter fiber clustering}

\name{\begin{tabular}[t]{c@{\extracolsep{1em}}c@{\extracolsep{1em}}c@{\extracolsep{1em}}c}
  \multicolumn{3}{c}{Narciso López-López$^{1,2}$, Andrea Vázquez$^{1}$,}\\
  \multicolumn{3}{c}{Cyril Poupon$^{3}$, Jean-Fran\c{c}ois Mangin$^{3}$ and Pamela Guevara$^{1}$}
  \thanks{This work has received funding by the CONICYT PFCHA/ DOCTORADO NACIONAL/2016-21160342, CONICYT FONDECYT 1190701, CONICYT PIA/Anillo de Investigaci\'on en Ciencia y Tecnolog\'ia ACT172121, Basal Center FB0008, and by the European Union’s Horizon 2020 research and innovation programme under the Marie Sklodowska-Curie grant agreement No 690941.}
\end{tabular}}

\address{$^{1}$ Universidad de Concepci\'on, Faculty of Engineering, Concepci\'on, Chile \\
$^{2}$Universidade da Coru\~na, LBD group, Dept. of Computer Science,\\ Facultade de Inform\'atica, CITIC, Campus de Elvi\~na, 15071, A Coru\~na, Spain \\
$^{3}$Neurospin, I2BM, CEA, Gif-sur-Yvette, France}

\begin{document}

\maketitle

\begin{abstract}
	
We present a hybrid method that performs the complete parcellation of the cerebral cortex of an individual, based on the connectivity information of the white matter fibers from a whole-brain tractography dataset. The method consists of five steps, first intra-subject clustering is performed on the brain tractography. The fibers that make up each cluster are then intersected with the cortical mesh and then filtered to discard outliers. In addition, the method resolves the overlapping between the different intersection regions (sub-parcels) throughout the cortex efficiently. Finally, a post-processing is done to achieve more uniform sub-parcels. 
The output is the complete labeling of cortical mesh vertices, representing the different cortex sub-parcels, with strong connections to other sub-parcels.
We evaluated our method with measures of brain connectivity such as functional segregation (clustering coefficient), functional integration (characteristic path length) and small-world. Results in five subjects from ARCHI database show a good individual cortical parcellation for each one, composed of about 200 sub-parcels per hemisphere and complying with these connectivity measures.

\end{abstract}


\section{Introduction}
\label{sec:intro}

Advances in brain imaging have allowed the study of the structure and connectivity of white matter (WM), a research area that is constantly growing. 
One of the most used techniques to understand the anatomical connectivity of the brain is the diffusion-weighted Magnetic Resonance Imaging (dMRI). 
It is a non-invasive and in-vivo technique, based on measurements of the movement of hydrogen molecules present in water \cite{le2015diffusion}. 
Tractography algorithms use dMRI information to estimate the main trajectories of the WM tracts \cite{mori2002fiber}. When applied to the whole-brain, resulting datasets contain a large amount of 3D polylines, called fibers, that represent the main brain WM connectivity.

Understanding how the brain works requires a detailed description of the network of connections that form it \cite{sporns2005human}.
A cortical parcellation represents a way to divide the brain cortex into macroscopic regions, according to their structure or functioning, in order to study brain connectivity \cite{de2013parcellation}. 
The best known parcellation is Brodmann's atlas, based on postmortem cytoarchitecture study, focused on the size, density, shape and distribution of cell bodies in cortical layers \cite{amunts2015architectonic}. In contrast, in-vivo techniques based on MRI, enable the development of other parcellations, based on anatomical structures \cite{destrieux2010automatic}, functional MRI (fMRI) \cite{schaefer2017local} or a multi-modal approach \cite{glasser2016multi}. Performing a cortical parcellation is a difficult task due to the high variability that exists between subjects in terms of white matter and gray matter, as well as the disadvantages of each imaging modality. 

The most common approaches to estimate brain connectivity are diffusion tractography, structural covariance, resting-state functional connectivity and meta-analytic connectivity modeling \cite{eickhoff2018imaging}.
Diffusion tractography provides information about structural connectivity, but has the limitations of not being able to delimit the beginnings and terminations of fiber bundles \cite{eickhoff2015connectivity} and produce false positives due to the large number of fibers that cross between the WM tracts. In addition, short association fiber connections can be lost due to the limited resolution of the tractographic methods \cite{maier2017challenge}. Two strategies can be used to perform diffusion-based cortical parcellations.
One approach first determines corresponding connections across subjects and then creates a parcellation according to the main connections in all the subjects. For example, a fiber bundle atlas of superficial WM connections was used to segment bundles in a group of subjects and get some consistent parcels for the 10 analyzed subjects \cite{cortex}. The difficulty here is to detect a representative set of the common connections for a population of subjects and create the final parcels. The second approach detects robust individual parcels from the whole tractography dataset, and then manages to find and delineate consistent parcels across subjects \cite{Lefranc2016}.


In this work, we propose a new hybrid method of individual cortical parcellation based on WM connectivity and intra-subject fiber clustering, with automatic parcel labeling. Our goal is to perform a good quality individual cortical parcellation to be used for a group-wise parcellation in the future. We applied the method to a group of subjects and evaluated several measures of brain connectivity. We demonstrate that the resulting networks for each subject comply with the integration and functional segregation as well as with the small-world definition.

\section{Materials and Methods}
\label{sec:matmet}

\subsection{Database and tractography datasets} 

We used the ARCHI database, 
composed of 79 healthy subjects \cite{Schmitt12}, and acquired with a 3T MRI scanner (Siemens, Erlangen). The MRI protocol included the acquisition of a T1-weighted dataset using an MPRAGE sequence (160 slices; matrix=256x240; voxel size=1x1x1.1 mm) and a SS-EPI single-shell HARDI dataset along 60 optimized DW directions, b=1500 $s/mm^2$ (70 slices, TH=1.7 mm, TE=93 ms, TR=14,000 ms, FA=90, matrix=128x128, RBW=1502 Hz/pixel, echospacing ES=0.75 ms, partial Fourier factor PF=6/8; GRAPPA = 2).

By using BrainVISA/Connectomist-2.0 software \cite{duclap2012connectomist}, data were pre-processed. The outliers were removed and the sources of artefacts were corrected. Next, to obtain ODF fields in each one of the voxels, the analytical Q-ball model was computed, and streamline deterministic tractography was performed on the entire T1-based brain mask, with a forward step of 0.2 mm and a maximum curvature angle of 30$^\circ$.
Moreover, to convert the data between the spaces Talairach, T1 and T2, the transformation matrices are available in the database.

\subsection{Diffusion-based cortical parcellation method}

\begin{figure*}[h] 
	\centering
	\includegraphics[width=16.0cm]{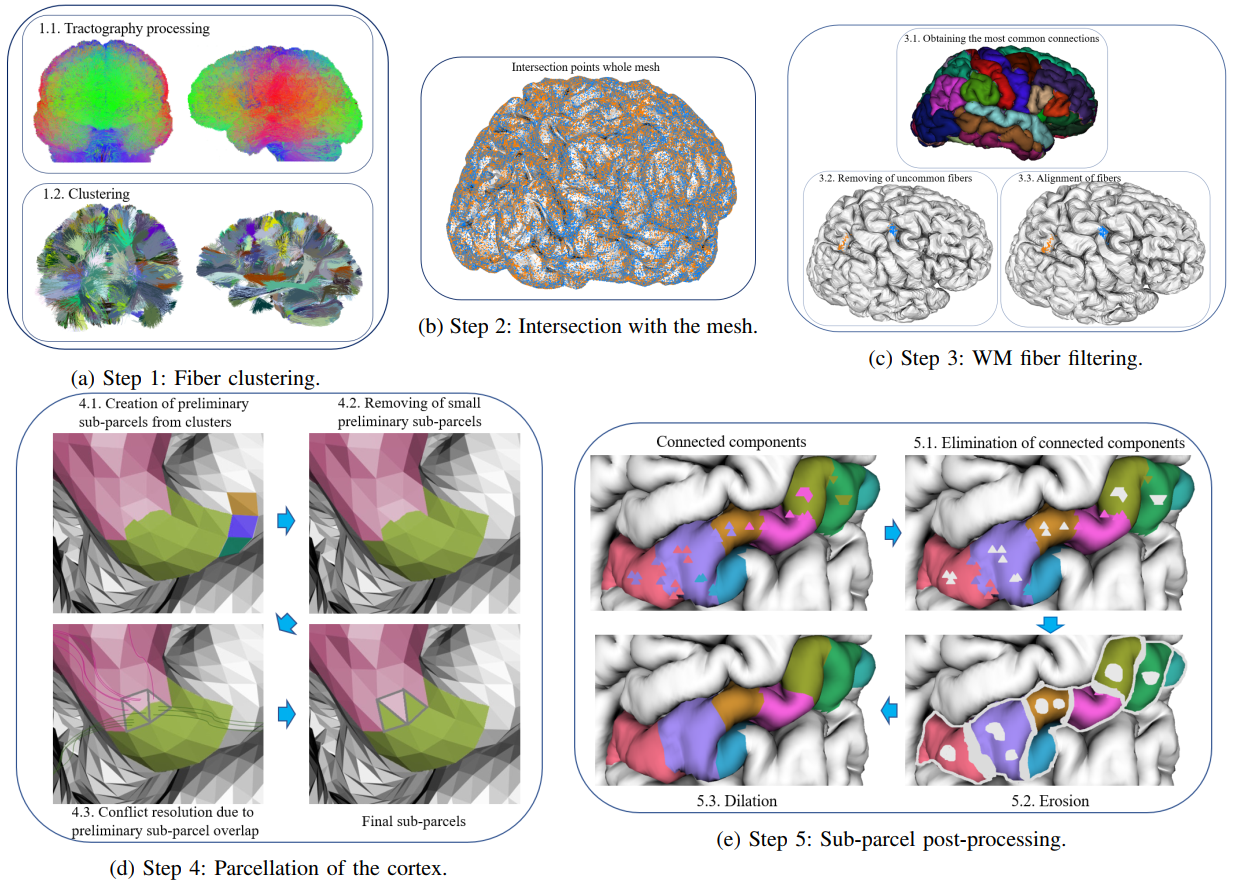}  
	\caption{Parcellation method. \textbf{Step 1: Fiber clustering}. First, the whole tractography is resampled with 21 points and transformed to T1 space. Next, a fiber clustering is applied to obtain compact clusters. \textbf{Step 2: Intersection with the mesh}. The intersection of the fiber clusters with the cortical mesh is calculated.  \textbf{Step 3: WM fiber filtering}. The fibers of each cluster are labeled according to an anatomical parcellation. Fibers that do not correspond to the most common connections are filtered out. Also, inverted fibers are realigned. \textbf{Step 4: Parcellation of the cortex}. Preliminary sub-parcels, from each cluster extremity are created. Next, small preliminary sub-parcels are removed. Finally, the overlap among sub-parcels is solved, by assigning each triangle to the most connected sub-parcel. \textbf{Step 5: Sub-parcel post-processing}. The main connected component of each sub-parcel is kept, in order to remove small isolated areas. Next, an erosion followed by a dilation (opening operation) are applied to eliminate some imperfections in the perimeter of the sub-parcels. }
	\label{fig:stepsParcellation}
\end{figure*}

We perform the parcellation of the cortex with a hybrid method based on WM connectivity given by fiber clusters, leading to an automatic labeling of cortical regions. In the following, we explain the whole method (see Figure \ref{fig:stepsParcellation}), which is composed of five steps: \textit{(1)} fiber clustering, \textit{(2)} intersection with the mesh, \textit{(3)} WM fiber filtering, \textit{(4)} parcellation of the cortex and \textit{(5)} sub-parcel post-processing.

\hfill\break
\textbf{STEP 1: Fiber clustering}: We apply an intra-subject clustering to all the fibers of the whole-brain tractography dataset. The objective is to create clusters with similar fibers, according to their position and shape, which we call fiber bundles. 

First, the tractography obtained from the database is preprocessed (see Figure \ref{fig:stepsParcellation}(a)$_{1.1}$). 
Fibers are resampled with 21 equidistant points, a number of points big enough to represent all the brain fibers. Next, the tractography datasets are transformed from T2 to T1 space.

Then, an intra-subject clustering algorithm is applied, which is composed of three main steps \cite{sanchezS}. First, a Minibatch k-means clustering \cite{sculley2010web} is performed in parallel over a subset of fiber points from all the fibers in the dataset. We choose minibatch k-means since it has low spatial and temporal complexities obtaining good quality of the clusters. The Elbow method \cite{kodinariya2013review} is used to determine the optimal number of clusters for k-means. Next, resulting point clusters are used to create fiber clusters by mapping, i. e. fibers with points sharing the same point clusters are grouped. This is constructed using a dictionary where each fiber has the cluster labels as a key, and the value in the dictionary is the group of fibers that share the same key. Finally, a merge is made of the final clusters that share the central point and are nearby. Next, for each group we calculate the maximum direct ($d_{E}$) and flipped ($d_{Ef}$) distance for all the fibers, and thus be able to compute the maximum Euclidean distance ($d_{ME}$ see equation~\ref{eq:dme}) between the relevant points. Equations \ref{eq:de} and \ref{eq:def} show the computation of the Euclidean distance $d_{E}$ and $d_{Ef}$, while equation~\ref{eq:dme} describes the maximum Euclidean distance:
\begin{equation}
\label{eq:de}
d_E(a,b) =||a-b||=(\sum_{i=1}^{k}(a_i-b_i)^2)^{1/2} 
\end{equation}
\begin{equation}
\label{eq:def}
d_{Ef}(a,b) = d_E(a,b^f)= d_E(a^f,b)
\end{equation}
\begin{equation}
\label{eq:dme}
d_{ME} = min(max(d_E(a,b)), max(d_{Ef}(a,b))) 
\end{equation}

where $a$ and $b$ are the coordinates of the fiber points in the 3D space, and $k$ $=$ 21 because it is a sufficient number of points to consider for obtaining a good result. We use the equation~\ref{eq:dme} since the orientation of the fibers obtained from the tractography is unknown. 
Finally, all the cluster centers that comply with $d_{ME}$ $<$ $d_{max}$ are merged. Figure \ref{fig:stepsParcellation}(a)$_{1.2}$ shows an example of the results after applying the clustering to a tractography dataset.

\hfill\break
\textbf{STEP 2: Intersection with the mesh}: This step determines the intersection of the fiber clusters with the cortical mesh, and was modified from \cite{silva2019cortical}. Figure \ref{fig:stepsParcellation}(b) shows an example of intersection points of the fibers with the mesh. 

First, a subdivision of 3D space into small cubic cells (around 1.5 mm) is implemented, to optimize spatial searches in the mesh. Then, some projection points are calculated from each fiber endpoint, detecting the cells and the neighborhood that touch the endpoints. Finally, the Möller-Trumbore algorithm \cite{moller2005fast} is used to calculate the intersection of the fibers and the mesh triangles. 
Equation~\ref{eq:Moller} shows the intersection algorithm:
\begin{equation}
\label{eq:Moller}
O +tD = (1-u-v)V_0 + uV_1 + vV_2
\end{equation}

where $O$ is the ray of origin, $t$ is the distance, $D$ is the normalized ray direction, $(u, v)$ are the coordinates of intersection in the triangle and $V_0$, $V_1$ and $V_2$ are the vertices of the mentioned triangle. The conditions for the coordinates are $u \ge 0$, $v \ge 0$ and $u +v \le 1$. Finally, the triangle intersects if there are permitted values within the ranges for the variables $t$, $u$ and $v$.

\hfill\break
\textbf{STEP 3: WM fiber filtering}: This step aims to label the clusters that intersect with the mesh and filter them to delimit anatomical cortical regions (brain circonvolutions). 

\begin{figure}[h] 
	\centering
	\includegraphics[width=8.5cm]{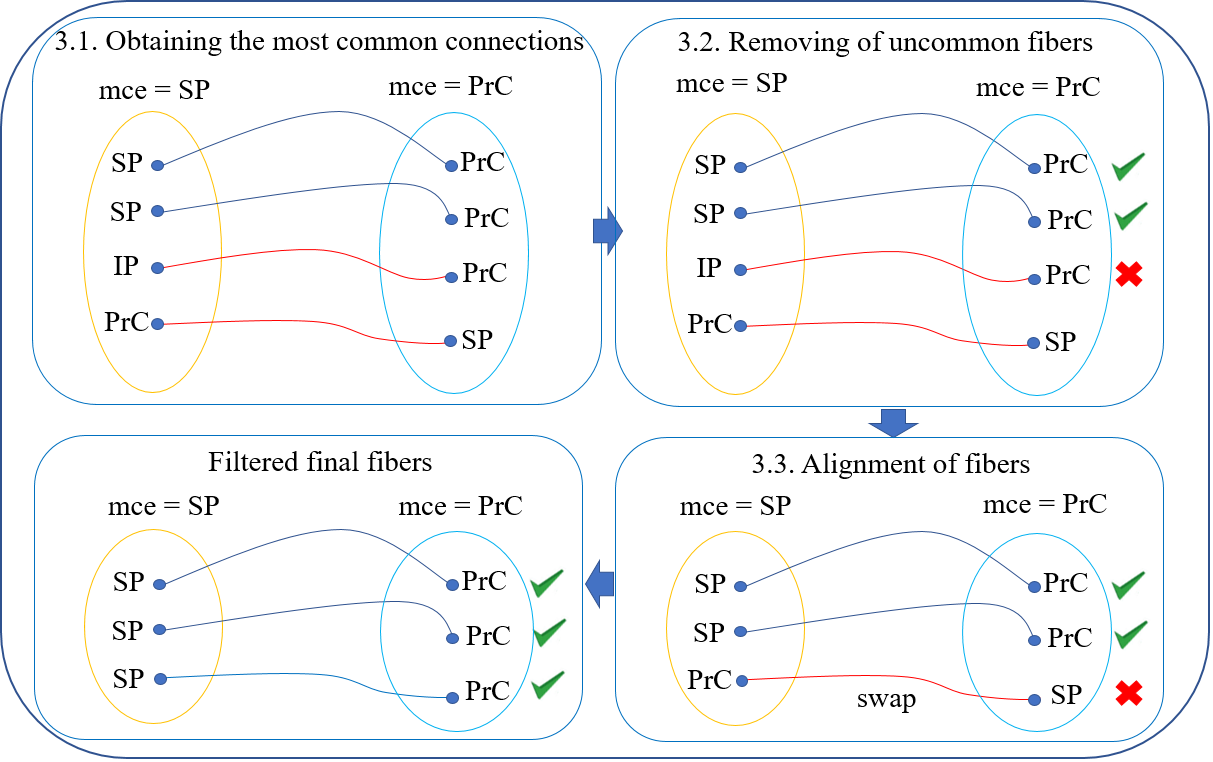}  
	\caption{Example of WM Fiber filtering for a cluster. \textbf{Sub-step 3.1}: Obtaining the most common connections. In this case, the most common connections are SP and PrC. \textbf{Sub-step 3.2}: Removing of uncommon fibers. The fiber labeled with the IP parcel is removed since it does not belong to SP. \textbf{Sub-step 3.3}: Alignment of fibers. Fibers that are inverted according to the most common connections, are swapped.}
	\label{fig:filter_mce}
\end{figure} 

To perform the filtering, we carry out three sub-steps (see Figure \ref{fig:stepsParcellation}(c)):
\begin{enumerate}
	\item \textbf{Obtaining the most common connections}:
	First, the fibers of each cluster are labeled according to the Desikan-Killiany cortical atlas labels \cite{desikan2006automated}. For that, we use the mesh vertex labeling information given by Freesurfer (see Figure \ref{fig:stepsParcellation}(c)$_{3.1}$). Next, the most common connection, at the beginning and the end of each cluster, are determined. Figure \ref{fig:filter_mce}$_{3.1}$ shows an example for different clusters connecting the Superior-Parietal (SP) and PreCentral (PrC) parcels.
	
	\item \textbf{Removing of uncommon fibers}:
	The next sub-step aims to remove the fibers that do not correspond to the most common connections, in order to keep only the fibers corresponding to the anatomical parcels given by the Desikan-Killiany cortical atlas. 
	See Figure \ref{fig:filter_mce}$_{3.2}$ for an example. 
	
	\item \textbf{Alignment of fibers}:
	Finally, a fiber alignment 
	is applied, where fibers that are inverted by respect to the most common connections are swapped, as shown in Figure \ref{fig:filter_mce}$_{3.3}$.
\end{enumerate} 

Once the aforementioned sub-steps have been carried out, white matter fiber clusters are filtered, getting a better delimitation between bundles.
  
\hfill\break
\textbf{STEP 4: Parcellation of the cortex}: This step creates sub-parcels from the preliminary sub-parcels defined by the intersection of each cluster. It  solves the conflicts between the overlaps of the preliminary sub-parcels. The creation of the sub-parcels is done in three sub-steps:
\begin{enumerate}
	\item \textbf{Creation of preliminary sub-parcels from clusters}:
	Each fiber cluster will define two preliminary sub-parcels, one from each extremity of the cluster. The set of preliminary sub-parcels is created by constructing a list of the triangles intersecting each cluster extremity 
	(see Figure \ref{fig:stepsParcellation}(d)$_{4.1}$). Note that a triangle can be intersected by several cluster extremities.
	
	\item \textbf{Removing of small preliminary sub-parcels}:
	Preliminary sub-parcels, that are too small, with a size 10\% smaller than the average size of the preliminary sub-parcels within an anatomical parcel, are removed (see Figure \ref{fig:stepsParcellation}(d)$_{4.2}$). With this step a big amount of isolated triangles and noisy preliminary sub-parcels are removed.
	
	\item \textbf{Conflict resolution due to preliminary sub-parcel overlap}:
	Some preliminary sub-parcels present an overlapping within an anatomical region. To solve this problem, the conflicting triangles (belonging to several preliminary sub-parcels) are analyzed, and assigned to the sub-parcel with the higher number of intersecting fibers (see Figure \ref{fig:stepsParcellation}(d)$_{4.3}$). 
\end{enumerate}

At the end of this step, the existing conflicts between preliminary sub-parcels disappear, thus obtaining a set of sub-parcels corresponding to each anatomical parcel (circonvolution).\\

\textbf{STEP 5: Sub-parcel post-processing}: This is the last step of the cortical parcellation method, that aims to perform a refinement of the sub-parcels, in order to get more uniform areas. It consists of three sub-steps:
\begin{enumerate}
	\item \textbf{Elimination of connected components}:
	Within an anatomical parcel, a graph is created for each sub-parcel and then, the connected components of each graph are calculated. The largest connected component of each sub-parcel is kept, leading to the removal of small isolated areas (see Figure \ref{fig:stepsParcellation}(e)$_{5.1}$).
	
	\item \textbf{Erosion}:
	The edges of each sub-parcel are eroded over the mesh, to finish the removal of some peaks or protrusions that deform the sub-parcel perimeter (see Figure \ref{fig:stepsParcellation}(e)$_{5.2}$).
	
	\item \textbf{Dilation}:
	Finally, the sub-parcels are expanded, filling the gaps left by erosion, resulting in smooth, uniform and well-defined sub-parcels (see Figure \ref{fig:stepsParcellation}(e)$_{5.3}$). The morphological operation of \textit{erosion + dilation} corresponds to an \textit{opening}, which is an operation for noise elimination.
\end{enumerate}

After carrying out the post-processing, we obtain the complete parcellation of the cortical mesh. It is defined by the subdivision of the cortex into sub-parcels, given by a label for each mesh vertex.


\section{Results}
\label{sec:res}
Almost all the parcellation steps were implemented in Python programming language (Python 3.6), with the exception of the intersection algorithm, made in C++11, which is parallelized with OpenMP. 
All the experiments ran on a computer with an 8-core Intel Core i7-6700K CPU running at 4GHz, 8MB of shared L3 cache and 8GB of RAM, using Ubuntu 18.04.2 LTS with kernel 4.15.0-55 (64 bits).

We calculated the cortical parcellation of five subjects from the ARCHI database, using their complete clustered tractographies. 
To evaluate the quality of the connections among parcels, we generated a connectivity map for each subject, from the resulting parcellation, and evaluated it using network graph metrics.
A connectivity map is built up by the tractography of each subject and the mesh parcellated into sub-parcels, performing the following steps: 
\begin{enumerate}
	\item The intersection of the complete tractography with the parcellated mesh is calculated.
	\item A square matrix $n*n$ is created with $n$ equal to the total number of sub-parcels, initialized to 0.
	\item For each fiber in the tractography, connecting two sub-parcels, a 1 is added to the cells corresponding to both sub-parcels in the connectivity matrix.
	
\end{enumerate}

Hence, a connectivity map was calculated for each subject, from its individual parcellation and tractography.\\

\subsection{Measures of brain connectivity}

There are many metrics for the evaluation of the characteristics of brain networks \cite{cohen2016segregation}. The properties we choose to analyze them are functional segregation (Clustering Coefficient), functional integration (Characteristic Path Length) and Small-World \cite{rubinov2010complex}. These properties have been shown to be present in the brains of the higher vertebrates \cite{tononi1994measure}:

\begin{enumerate}
	\item Functional segregation:
	It is the presence of strongly interconnected groups or clusters in the brain. The metric used to measure this property is the Clustering Coefficient \cite{watts1998collective}. A value closer to 0 denotes a random network, however, a complex-network shows higher clustering coefficient values. Equation~\ref{eq:cluscoef} defines the clustering coefficient for undirected graphs:
	\begin{equation}
	\label{eq:cluscoef}
	C_i = \frac{2N_i}{k_i(k_i-1)}  
	\end{equation}
	
	where $N_i$ is the amount of links in the neighborhood of $i$, $k_i$ is the degree of a particular node $i$ and $C_i$ is the clustering coefficient for node $i$.
	
	The equation~\ref{eq:cluscoefavg} measures the average clustering coefficient for the entire network:
	\begin{equation}
	\label{eq:cluscoefavg}
	C = \frac{1}{n} \sum_{i\in G} C_i
	\end{equation}
	
	where $G$ is the graph of the undirected network, $n$ is the total number of nodes in the network and $C$ is the average clustering coefficient.
	
	\item Functional integration:
	It is the ability to easily distribute information across the different specialized regions of the brain. The better the information is distributed, the higher is the functional integration. The measure used to measure this property is the Path Length, specifically the Characteristic Path Length  \cite{watts1998collective} that averages the shortest path length between each pair of nodes (sub-parcels) in the network.
	Equation~\ref{eq:pathavg} describes the characteristic path length for an undirected graph:
	\begin{equation}
	\label{eq:pathavg}
	L = \frac{1}{n(n-1)} \sum_{i,j\in G, {i\neq j}} d_{ij}
	\end{equation}
	
	where $G$ is the graph of the whole undirected network, $n$ is the total number of nodes in the network, $d_{ij}$ between nodes $i$ and $j$ is the smallest distance between them and $L$ is the characteristic path length.
	
	\item Small-World: 
	This metric is very relevant, since it combines the two previous ones. A brain network must have good functional segregation, keeping functional integration a little lower, that is, strongly interconnected internal regions, and in turn, a good amount of links to other regions \cite{rubinov2010complex}. The $\sigma$ coefficient is used to measure the property of small-world, which is the ratio between the clustering coefficient and its equivalent random network divided by the path length and its corresponding random network. Equation~\ref{eq:swsigma} details the sigma coefficient:
	\begin{equation}
	\label{eq:swsigma}
	\sigma = \frac{\frac{C}{C_r}}{\frac{L}{L_r}}
	\end{equation}
	
	where $C$ is the clustering coefficient, $C_r$ is the clustering coefficient for the equivalent random network, $L$ is the path length, $L_r$ is the path length of its equivalent random network, and $\sigma$ is the coefficient that measures the small-world. A network is considered small-world if $C$ $\gg$ $C_r$ and $L$ $\approx$ $L_r$, then $\sigma$ $>$ 1.

\end{enumerate}

\begin{figure}[h] 
	\centering
	\includegraphics[width=9.0cm]{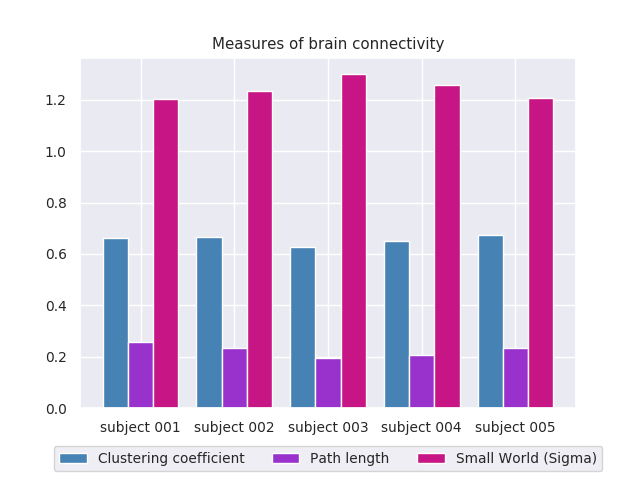}  
	\caption{Measures of brain connectivity obtained for the five subjects: Clustering Coefficient, Characteristic Path Length and Small-World.}
	\label{fig:binary_multiple_barchar}
\end{figure}

To evaluate the network we used the \textit{bctpy} toolbox\footnote{https://github.com/aestrivex/bctpy} for Python, that provides functions to calculate the Clustering Coefficient and the Characteristic Path Length metrics. Moreover, to calculate the Small-World metric, we transformed our matrices into graphs and used the \textit{networkx} library for Python \cite{humphries2005brainstem}. Figure \ref{fig:binary_multiple_barchar} shows the metrics for the five subjects. A high Clustering Coefficient, while maintaining a lower Characteristic Path Length, and therefore a Small-World $>$ 1 was obtained for all the subjects.

These results demonstrate that the connectivity maps obtained for each subject, created by our parcellation method, are considered small-world networks, and therefore, maintain the properties of segregation and functional integration of the brain. In addition, as seen in Figure \ref{fig:binary_multiple_barchar}, the results are very similar among the five subjects.

\subsection{Qualitative Analysis}

In this subsection we show the different views of the cerebral cortex of a subject, as well as the different individual parcellations for five subjects in the database.

\begin{figure}[t!] 
	\centering
	\includegraphics[width=8.5cm]{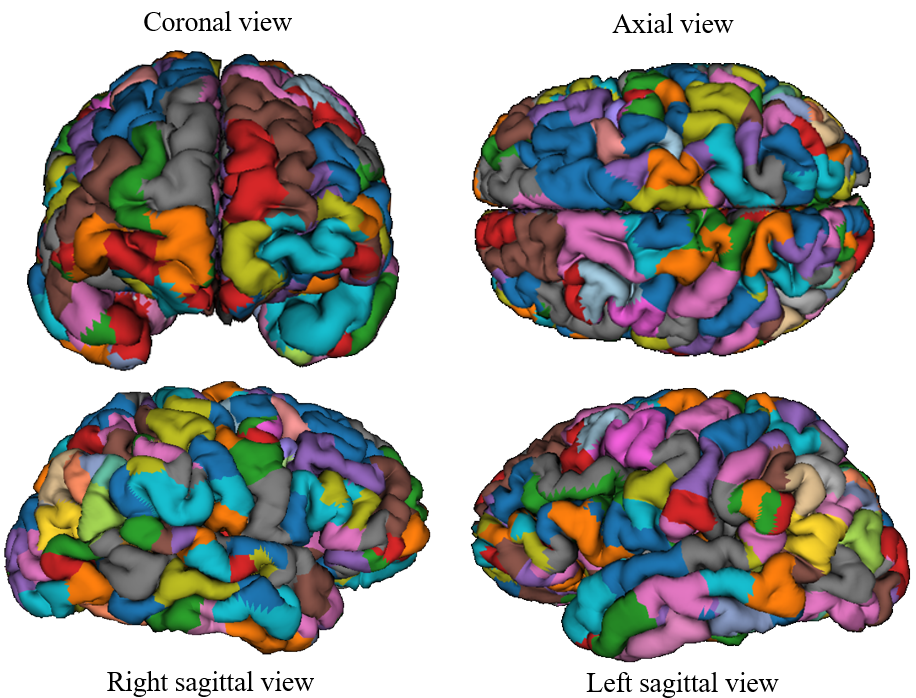}  
	\caption{Individual parcellation for subject 001. Coronal, axial, right and left sagittal views are displayed. The parcellation subdivides the cortex into 430 sub-parcels, 209 in the left hemisphere and 221 in the right hemisphere.}
	\label{fig:allviews}
\end{figure}

\begin{figure}[t!] 
	\centering
	\includegraphics[width=8.5cm]{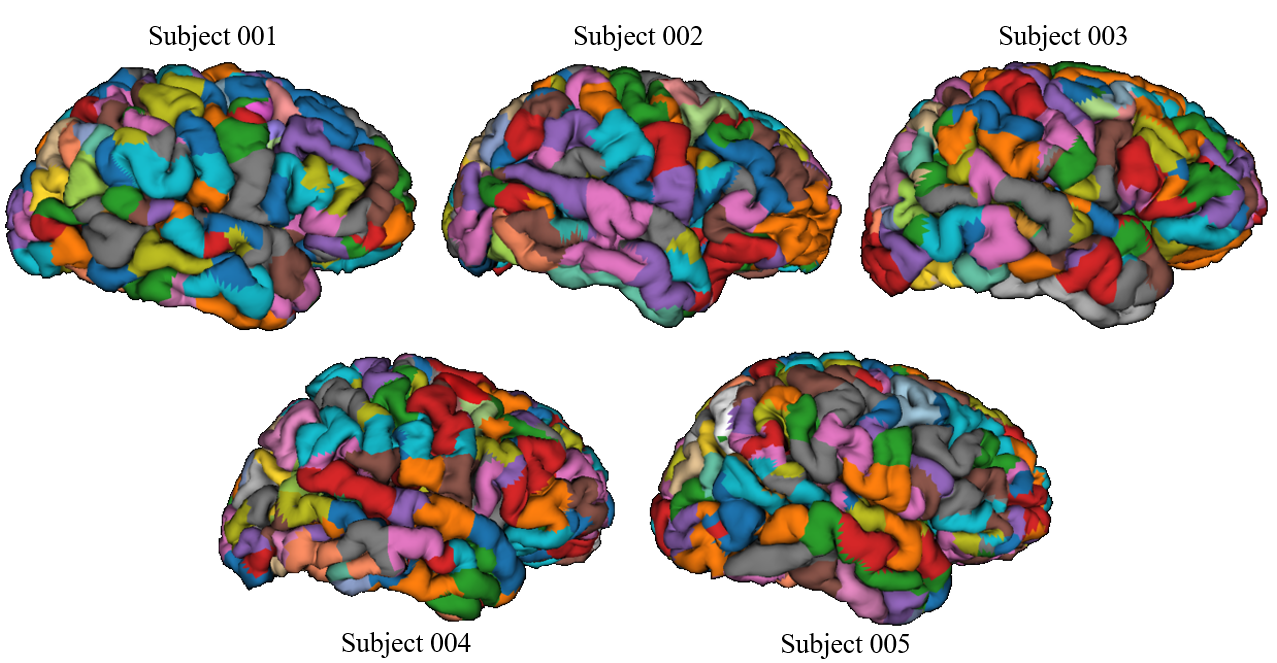}  
	\caption{Individual cortex parcellation for five subjects (right sagittal views). The average of sub-parcels obtained for a hemisphere is 200, with 400 sub-parcels on average of the entire cerebral cortex.}
	\label{fig:subjects}
\end{figure}

Figure \ref{fig:allviews} shows the coronal, axial, right and left sagittal views, resulting from the individual parcellation of subject 001, consisting of 430 sub-parcels for the whole brain, with 209 in the left hemisphere and 221 in the right hemisphere. Finally, Figure \ref{fig:subjects} displays the parcellation results for the five subjects. An average of 400 sub-parcels was obtained for the whole cortex, with approximately 200 sub-parcels in average per hemisphere.

\section{Conclusions}
\label{sec:con}

We developed a hybrid method for the individual cortex parcellation, based on the connectivity of WM fiber clusters. The fiber clustering helps to define compact connections and filter out outliers. The method provides good quality results in the connectivity maps of the five analyzed subjects, evaluated by network graph metrics. Resulting networks show a high Clustering Coefficient, low Characteristic Path Length and Small-World property. These properties indicate good integration and functional segregation of the brain \cite{rubinov2010complex}.
As future work, 
we will explore the implementation of a multi-subject version of this parcellation method and test it in different databases, such as the Human Connectome Project. Hence, we could obtain an atlas (or model) of  cortical parcels with similar connectivity profiles across a population of healthy subjects. 
Also, other information could be integrated, like fibers segmented with a bundle atlas \cite{vazquez2019parallel}, or data from other modalities, like fMRI.


\bibliographystyle{IEEEbib}
{\small \bibliography{strings,refs}}

\end{document}